\documentstyle[prd,aps,preprint,epsf,floats]{revtex}
\draft
\tighten

\newcommand{\cfig}[2]{\hbox to \columnwidth {\hfil\epsfysize = #2 \epsfbox{#1}
\hfil}}

\begin{document}

\preprint{
         \parbox{1.5in}{%
           PSU/TH/183
         }
}

\title{Formation of Spiral Structures in Galactic Discs from 
Reaction-Diffusion Networks}

\author{Andreas Freund$^{a}$\footnote{e-mail:freund@phys.psu.edu}}

\address{ $^{a}$Department of Physics, Penn State University,
               104 Davey Lab.,University Park, PA 16802, U.S.A.}

\maketitle

\begin{abstract}
    In this paper, we propose an extension of the Smolin-model of a galactic
    disc in isolated, flocculent Sc-type galaxies. This model supplies 
    the necessary mechanism to suppress spatial inhomogeneities on short 
    distance scales  not present in the model proposed by Smolin. 
\end{abstract}


\section{Introduction}
\label{sec:intro}

In astronomy, the problems of pattern formation appear on a broad variety 
of scales; from the formation of galaxy clusters on the scale of 50 to 
several hundred mpc down to the formation of stars on a scale
of a few dozen pc. The most intensely studied of these is the problem 
of the formation of spiral structures in galactic discs, which has brought
forth numerous theoretical and numerical approaches \cite{1,2,3,4,5,6,7,8},
which produce impressive results, but also have their own shortcomings. 
Models based on propagating starformation
\cite{2} do indeed reproduce the appearance of a range of galaxies \cite{4,5},
however, they either involve strong simplifications of the astronomy involved 
or need the fine tuning of rates in order to reproduce the observed structures.
In the case of density wave models \cite{3} where density waves facilitate 
the star formation process, 
a continous recurrance of density waves passing through the galaxy is required
in order to keep the star formation process continous over the galaxies 
lifetime, hence there is the need for sufficient sources of density waves 
requiring large massive objects, like other galaxies, not being too distant 
from the galaxy in question. However, if one is concerned with isolated 
galaxies where there cannot be a continous recurrance of density waves, 
there has to be another process or processes to produce the observed spiral 
structure other than density waves.  

As a consequence of the above, in the following, we will be concerned 
with flocculent, spiral structure in so called Sc-type galaxies, which 
are sufficiently far from other galaxies, such that the spiral structure 
has to be endogenous, i.e. understood as a product of processes occuring 
within the disc. As was pointed out by the authors of Ref.\ \cite{4,6} these 
galaxies show spiral structure in the blue light but not in the red light; 
hence the structure is mainly not due to a density wave but is rather a 
trace of a star formation process \cite{1}.

It is important to point out that star formation apparently occurs at a 
constant rate in these galaxies when averaged over the whole disc. This is an
important clue, as pointed out by Smolin \cite{1}, since the implication 
of this constant rate is the presence of feedback mechanisms in the processes
governing the rate of star formation, so as to keep the rate steady and slow
\cite{10,11}. Since the rate is constant over the history of a galaxy 
($10^{10}$ years) as compared to the time scales of the dynamical process 
($10^7$ years), there is no other known way to explain that galaxies with slow 
and steady rates of star formation are so common. Hence, one can put forth the 
hypothesis \cite{1} that the star formation process can be explained to be 
a network of self-regulated and autocatalyzed processes resulting from the 
self-organization of the material in the disc. Thus non-equilibrium 
statistics govern the evolution of the disc system in which the 
network
processes, involving matter- and energy-flows amongst its components, are
governed by feedback-loops. In other words, the way spatial inhomogeneities
are created and stabilized from non-equilibrium systems might explain the 
patterns produced by the star formation process.

There are numerous theoretical and experimental articles (the 
interested reader should consult the references contained in \cite{1}) 
describing how spatial and temporal patterns are produced in non-equilibrium 
systems. 
However, we would like to point out one important aspect. Recent studies of 
non-equilibrium systems have produced impressive successes by which patterns,
that can be reproduced in actual laboratory situations, are explained by
simple models involving both partial differential equations and discrete 
elements like cellular automata. Hence, structure formation seems to occur in
a broad variety of far-from-equilibrium steady-state systems. To see whether 
this is true in the above mentioned case of flocculent galaxies let us give
a 'check-list' of the main elements that characterize systems to which 
non-equilibrium models of pattern formation are applicable. These include 
\cite{13,14,15,16,17,18,19,20,21}:
\begin{itemize}

\item The system is in a steady-state with a slow and steady flow of energy 
      (and matter) through it.

\item The steady-state is far from thermodynamic equilibrium i.e a 
      coexistence of several phases of matter exchanging matter and energy
      through closed cycles.

\item The flow-rate of material around these cycles is governed by 
      feedback-loops that have arisen during the organization of the system 
      to the steady state.

\item The reaction networks are autocatalytic i.e. substances serving as 
      catalysts/inhibitors of the reaction network are produced by 
      reactions within the network.

\item The possibilty of spatial segregation of different phases/materials in
      the network if catalysts/inhibitors, or more precisely their effects,
      propagate over different scales.

\end{itemize}         

Given these characteristics, there are a variety of models, particularly of the
reaction-diffusion type \cite{13,14,15} which describe how spatial structure
is formed and stabilized. It was shown in \cite{1} that the above mentioned 
criteria are met by the Sc-type galaxies and hence, the galactic disc of these
galaxies can be described as an autocatalytic network of reactions or put 
differently, the criteria, allowing the system to be described by a 
reaction-diffusion model, have been met.

In Sec.\ \ref{onezone} the main processes are reviewed drawing heavily on the 
excellent exposition in \cite{1}, a homogenous ``one
zone model'' very close to the Smolin-model \cite{1} is set-up, and it is 
shown that the system reaches a steady state.
In Sec.\ \ref{reacdif} the model is expanded to allow for spatial variations
\footnote{The motivation for using diffusion was taken from \cite{1}.}
In Sec.\ \ref{linana} a linearized analysis is carried out and a closer look 
is taken at the region in parameter space for which spatial inhomogeneities
occur in the model. It will be shown that the model proposed in this paper 
remedies one mayor drawback of the Smolin-model, namely the non-suppresion of
instabilities at small scales. Furthermore there are at least two distinct 
regions
in the parameter space which allow for qualitatively different results in the 
sensitivity of the model to changes in the main parameters. Conclusions and 
future work will be discussed in Sec.\ \ref{concl}.


\section{The homogenous ``one zone model''}
\label{onezone}

Before describing the ``one zone model'', let us briefly review the major 
processes involved in the model.

\begin{itemize}

\item {\bf Condensation of giant molecular clouds(GMC's)}

      GMC's are cold clouds condensing out of the interstellar medium (ISM),
      apparently forming scale invariant or fractal distributions of cold 
      molecular gas and dust.

      \begin{itemize}

      \item Catalysts: Dust, carbon and oxygen.\footnote{
Note that these substances spread through the ISM over an intermediate
distance scale $L_{int}\simeq 100$pc, corresponding to the scale over which
the products of massive stars and supernovae are spread due to the 
supernovae's shockwaves.
}

      \item Inhibitors: The two main inhibitors are ultraviolet radiation from
            massive stars which heats the ambient ISM making condensation 
            less likely\footnote{
This useful picture was introduced by Pavarrano et al. \cite{11}.
}
            and shockwaves from supernovae dispersing the GMC's. UV radiation 
            is inhibitory on a very long distance scale ($L_{long}\simeq$ 
            size of galaxy) with a very short propagation time of about $10^5$
            years, whereas the inhibitory effect of the shockwaves is on a 
            short distance scale ($L_{short}\simeq 20$pc) due to the fact that
            the ISM's viscosity, with respect to shockwaves, slows these 
            waves down, enough to be ineffective as an inhibitor on 
            $L_{int}\simeq 100$pc.
      \end{itemize}

\item {\bf Collapse of GMC's}

      Star formation occurs when the cores of GMC's collapse. There is a small
      possibility for a spontaneous rate of core collpase, however the 
      collapses of cores massive enough to lead to formation of massive 
      stars is usually catalyzed.

      \begin{itemize}
   
      \item Catalysts: The main catalyst are shockwaves from supernovae or
            HII-regions. These processes have a typical scale of 
            $L_{int}$.\footnote{
Density waves can also cause core collapses. These are probably dominant in
Grand Design Spirals but are of lesser importance for Sc-type galaxies
}

      \item Inhibitors: The main inhibitors are stellar winds from young,
            massive objects disrupting GMC's in which they are formed as
            well as UV radiation from young, massive stars, 
            evaporating  GMC's. The last process is \underline{not directly}
            related to the heating of the ISM. The typical distance scales 
            are $L_{short}\simeq$ size of one cloud complex.
      \end{itemize}        
      
      The effect of collapsing GMC's is the introduction of a latency time
      $\tau_{L}$ during which the star forming process will not recur. 
      $\tau_{L}$ is of the order of the time it takes for a GMC to form out 
      of gas, after another GMC in the same region has been evaporated as 
      a result of radiation from newly formed stars.

\item {\bf Star Formation}

      The final stages of star formation involve the formation of a protostar
      and an accretion disc. The process is self-limiting (see \cite{12}) i.e.
      matter accretes onto the protostar until it is stopped by the outflow 
      of matter due to stellar winds created in the accretion disc after the
      start of nuclear reactions.
 
\end{itemize}

Let us move on and start discussing the homogeneous ``one zone model''. 
Neglecting spatial variations for the moment, the reaction network may be 
described by a ``one zone model'' analogous to a system of equations describing
homogeneous, chemical reaction networks. Labeling the densities of components
as:
\bigskip

{\hskip 2in $c$ = cold gas in GMC's}

{\hskip 2in $w$ = warm, ambient gas}

{\hskip 2in $s$ = massive stars}

{\hskip 2in $d$ = light stars}

{\hskip 2in $r$ = density of UV radiation}

{\hskip 2in $h$ = density of shockwaves from supernovae}

\bigskip
we get, following the spirit of the chemical reaction models:
\begin{eqnarray}
\dot c &=& \frac{\alpha 'w^2}{r} -(\beta + \mu )ch - \gamma cs - \gamma 'cr
\nonumber\\
\dot s &=&\beta ch - \frac{s}{\tau}\nonumber\\
\dot w &=&- \frac{\alpha 'w^2}{r} + \frac{s}{\tau} + \delta + \gamma 'cr +
\gamma cs\nonumber\\
\dot r &=&- \phi (c + w)r + \eta s\nonumber\\
\dot h &=&- \phi '(c + w)h + \eta 's\nonumber\\
\dot d &=&\mu ch.
\label{eq.staedy}  
\end{eqnarray}
The parameters describe the rates of different, astrophysical processes:

\bigskip
{\hskip 0.5in
        $\tau$ = life time of a typical massive star ($10^7$ years).\footnote{
$\tau$ will also govern, as it turns out, the characteristic time scales of
the instabilities. It is important for this scale not to be too long or too 
short such that the instabilities do neither grow too fast which would be in 
contradiction to the observed steady star formation rate nor too slow 
which too would not be in accord with the observation of a {\it steady}
star formation rate. This means that a scale of the order of the dynamical 
time scale of the system will be appropriate.}}

{\hskip 0.5in $\alpha '\sim$ characteristic rate for GMC's to condense
              from warm
              gas.\footnote{The efficiency for the production of massive 
              stars is small, hence one must choose parameters such that 
              $\frac{\alpha 'w}{r}>\tau^{-1}$ in the steady state.}}

\bigskip
$\beta$, $\mu$, $\gamma$ and $\gamma '$ govern the rates per unit mass 
density at which material flows from GMC's to other states of the ISM due to 
processes catalyzed by the action of shockwaves, UV radiation and massive 
stars.
\bigskip

{\hskip 0.5in $\beta$ = formation rate of massive stars per unit density of c.}

{\hskip 0.5in    $\mu$ = formation rate of light stars per unit density of c.}

{\hskip 0.5in $\gamma$ = rate per unit density that cold gas is heated by 
                       massive stars.\footnote{
                                   This is a a short scale/local effect.
}}

{\hskip 0.5in $\gamma '$ = rate per unit density that cold gas is heated by UV 
                        radiation.}

{\hskip 0.5in $\delta$ = rate at which warm gas accretes onto the galactic disc.}

{\hskip 0.5in $\phi$,$\phi '$ = rate at which radiation and shockwaves are 
                             damped by the ISM.}

{\hskip 0.5in $\eta ',\eta$ = rate at which shockwaves and radiation are 
                              produced by massive stars.}

\bigskip
The simplifications in the physics are the same as those made by Smolin 
\cite{1} in his model. Let us quickly recount: there are only two types of 
stars, light
stars can evaporate spontaneously from GMC's, the role of dust and carbon is
not explicitly included, the Pavarrano process is given by a simple negative
feedback and $\delta$ is assumed to be constant. Notwithstanding these grains 
of salt, the model is an acceptable first model of how the ISM arrives at a
steady-state. Solving Eq.\ \ref{eq.staedy} for the steady state (all time 
derivatives vanish except $\dot d$) one finds a unique solution:
\begin{eqnarray}
s_{0}&=& \beta \tau \frac{\delta}{\mu}\nonumber\\
h_{0}&=& \frac{\delta}{\mu c_{0}}\nonumber\\
r_{0}&=& \frac{\eta \phi '}{\eta '\phi}h_{0}\nonumber\\
w_{0}&=& \left [ \frac{\beta \tau \eta '}{\phi '} - 1 \right ]c_{0}\nonumber\\
0 &=&\frac{\mu^2 \eta ' \alpha '}{\delta^2 \eta}\left [1 + \beta \tau \eta '
\right ]
c_{0}^3 - \gamma \beta \tau c_{0} - \left [ \beta + \frac{\gamma ' \eta}
{\eta '}\right ],
\end{eqnarray}
also
\begin{equation}
\dot d = \delta,
\end{equation}
so that the steady-state forms light stars at the rate $\delta$. This, of 
course, has to be true since matter is conserved in the system!

 
\section{The reaction-diffusion system}
\label{reacdif}

The formalism of temporal and spatial inhomogeneities can be naturally 
understood in the realm of biology and chemistry, if autocatalytic 
reaction networks have catalysts/inhibitors spread through the system over 
different scales in space and time i.e. one has a hierarchy of scales over 
which catalytic/inhibitory reactions are alternatively more important (see
Ref.\ \cite{13,14,15,16,17,18,19,20,25} for details). The main claim of 
Ref.\ \cite{1} and of this paper is that this may give a natural understanding 
about the appearance of structures in galactic discs. As was pointed out in 
Sec.\ \ref{onezone} the above mentioned conditions are met (see \cite{1}).
The processes over $L_{int}$ are all catalytic and over $L_{short}$ and 
$L_{long}$ inhibitory.

Now that one has met the conditions for pattern formation in the model, one 
has to incorporate spatial inhomogeneities. This is most commonly done through
diffusion. Diffusion definitely occurs in biological systems, shockwaves and 
radiation however are probably best described by propagation or some 
Navier-Stokes type of flow-equation. Nevertheless the assumption of 
diffusion is a rather good first approximation, taking into account the 
rather impressive results of the Gerola-Seiden-Schulman model \cite{4} based
on cellular automata and the Elmegreen-Thomasson model \cite{5} where
shockwaves responsible for propagating star formation are indirectly modeled
as either a direct effect on the catalyzation of star formation in 
neighbouring regions or a process by which ``young stars'' diffuse from their
parent cloud and then give energy to a nearby GMC. In order to include 
diffusion in our equations we add a diffusion term $D\nabla^2$. 
We also simplify by normalizing all densities according to:
\begin{equation}
\bar c(x,t) = \frac{c(x,t)}{c_{0}}\ \ \mbox{etc.}   
\end{equation}
We then obtain the following system of equations:
\begin{eqnarray}
\dot{\bar c} &=& \alpha \frac{{\bar w}^2}{\bar r} + \left [ \nu + \frac{\rho}
{\epsilon\tau} + \frac{1}{\epsilon T} - \alpha \right ]\bar c \bar s - 
\nu\bar c\bar r - \left [ \frac{\rho}{\epsilon\tau} + \frac{1}{\epsilon T}
\right ]\bar c\bar h\nonumber\\
\dot {\bar s} &=& D_s\nabla^2\bar s + \frac{1}{\tau}\left[ \bar c\bar h - 
\bar s \right ]\nonumber\\
\dot {\bar w} &=& \alpha\epsilon\left [ \bar c\bar s - \frac{{\bar w}^2}
{\bar r} \right ] + \epsilon\nu\left [ \bar c\bar r - \bar c\bar s\right ] +
\frac{\rho}{\tau}\left [ \bar s - \bar c\bar s\right ] + \frac{1}{T}\left [
1 - \bar c\bar s\right ]\nonumber\\
\dot{\bar r} &=& D_r\nabla^2\bar r - \frac{\sigma\epsilon}{1 + \epsilon}
\bar c\bar r - \frac{\sigma}{1 + \epsilon}\bar w\bar r + \sigma\bar s
\nonumber\\
\dot {\bar h} &=& D_h\nabla^2\bar h - \frac{\sigma '\epsilon}{1 + \epsilon}\bar
 c\bar h - \frac{\sigma '}{1 + \epsilon}\bar w\bar h + \sigma '\bar s
\nonumber\\
\dot d &=& \delta\bar c\bar h.
\label{eq.rd}
\end{eqnarray}
with the following parameters:
\begin{eqnarray}
&\alpha = \frac{\alpha 'w_{0}^2}{c_{0}r_{0}}&\ \ ;\ \ \frac{1}{T} = 
\frac{\delta}{w_{0}}\nonumber\\
&\rho = \frac{s_0}{w_0}&\ \ ;\ \ \nu = r_0\gamma '\nonumber\\
&\epsilon = \frac{c_0}{w_0}&\ \ ;\ \ \sigma = \eta\frac{s_0}{r_0}\nonumber\\
&\sigma ' = \eta '\frac{s_0}{h_0}&.
\end{eqnarray}
where $\alpha$ gives the time of condensation of cold clouds, which can be 
taken to be about $10-100$ ($\tau = 10^7$ years, the typical time scale of 
our problem, is set to be $1$), $\rho$ is the ratio of mass in massive stars to
mass in warm gas and can be set to $0.1$ indicating that the efficiency of
the star formation process is low and the massive-star production is
suppressed by the power law in the initial mass function. $\epsilon$ is the 
ratio of cold to warm gas and should be around unity since about $50\%$ of
the gas in the ISM is observed to be in GMC's. $T^{-1}$ is the time scale for
accretion of warm gas onto the disc and should be about $100-1000$ times 
longer than $\tau$ since one needs a continous supply of matter over the 
galaxies
lifetime. $\sigma$ and $\sigma '$ are the parameters governing the
``viscosity'' of the ISM with respect to UV radiation and shockwaves and can
be taken to be about $10$ and $0.1$ respectively, again with $\tau = 1$,
since $\sigma^{(')} \sim \frac{\tau}{T'}$ where $T'$ (time of travel to the
typical scale of the process) can be reasonably taken to be $10^6$ and 
$10^8$ years respectively. $\nu$ is the parameter governing the 
propagation of the core collapsing effect of UV radiation and massive stars, 
taken to be $10^2$ to $10^3$ in units of $\tau$, since the 
effect takes about $10^5$ years to propagate through the disc\footnote{Here 
the main effect comes from UV-radiation which travels with the speed of light.}
. 
The diffusion constants $D_s,
D_r$ and $D_h$ can, reasonably, be taken to be $D_s=L_{int}^2/\tau = 10^{-3}$,
$10^5\leq D_r \leq 10^6$\footnote{ This is about the size of the radius of 
the galactic disc times the speed of light, since UV radiation propagates 
through the whole disc.} and $D_h \simeq 10$ \footnote{ This is about 
$L_{int}\times$ average
speed of shockwave $O(10^5\frac{m}{s})$.} in units of $\tau$. The greatest 
uncertainties lie in $\nu ,\sigma ,\sigma '$ and 
the diffusion constants since there is no data available on these parameters, 
i.e one has to check the sensitivity of the model 
to changes in these parameters.

As a next step, we can assume the problem to be essentially two dimensional
due to the fact that the disc is very ``thin'' as compared to its radius. One 
has thus arrived at a reaction-diffusion model governing the dynamics of the 
ISM. It has to be pointed out here, that the differential rotation of the disc 
is not included at this stage. It will be included in the full numerical 
analysis which will be the next stage of the project.
      
\section{Linearized analysis and the region in parameter space}
\label{linana}

In the following, the linearized analysis of the model proposed in Sec.\ 
\ref{reacdif} will be carried out to see if there are indeed unstable modes
which may develop into spatial structure. To do this, one expands Eq.\ 
\ref{eq.rd} to linear order from the steady-state:
\begin{equation}
\bar c = 1 + C\ \ \mbox{etc.},
\end{equation}
and arrives at the following system of equations:
\begin{eqnarray}
\dot C &=& \alpha\left[ 2W - C - S\right ] - \nu\left [ R - S\right ] 
-\left [ \frac{\rho}
{\epsilon\tau} + \frac{1}{\epsilon T}\right ] \left [ H - S\right ]\nonumber\\
\dot S &=& D_{s}\nabla^2 S + \frac{1}{\tau}\left [ C + H - S\right ]\nonumber\\
\dot W &=& \alpha\epsilon\left [ -2W + C + S\right ] + \epsilon\nu\left [ R - S
\right ] - \left [ \frac{\rho}{\tau} + \frac{1}{T}\right ]C - \frac{1}{T}S
\nonumber\\
\dot R &=& D_{r}\nabla^2 R - \frac{\sigma}{1 + \epsilon}\left [ \epsilon C + W
\right ] - \sigma\left [ R - S\right ]\nonumber\\
\dot H &=& D_{h}\nabla^2 H - \frac{\sigma '}{1 + \epsilon}\left [ \epsilon C 
+ W\right ] - \sigma '\left [ H - S\right ]\nonumber\\  
\dot d &=& \delta\left [ 1 + C + H\right ]
\label{eq.lin}
\end{eqnarray}

As an ansatz, one can take the following solutions to Eq.\ \ref{eq.lin}:
\begin{eqnarray}
C &=& \tilde C e^{\lambda t}\cos (k\cdot x)\nonumber\\
S &=& \tilde S e^{\lambda t}\cos (k\cdot x)\nonumber\\
W &=& \tilde W e^{\lambda t}\cos (k\cdot x)\nonumber\\
R &=& \tilde R e^{\lambda t}\cos (k\cdot x)\nonumber\\
H &=& \tilde H e^{\lambda t}\cos (k\cdot x)
\label{eq.ansatz}
\end{eqnarray}
which describe an instability with wavevector $k$ growing exponentially with a
time scale $\lambda^{-1}$. To discover instabilities, one has to find 
solutions for reasonable values of the parameters and wavelengths for which 
$\lambda$ is real and positive. Plugging Eq.\ \ref{eq.ansatz} into Eq.\ 
\ref{eq.lin}; one arrives at an eigenvalue problem $M^a_b v^b = \lambda v^a$
with $v^a = (C,S,W,R,H)$ and the matrix:
\begin{equation}
\pmatrix 
{-\alpha & -\alpha + \nu + \frac{\rho}{\epsilon\tau} + \frac{1}{\epsilon T} & 2
\alpha & -\nu & -\frac{\rho}{\epsilon\tau} - \frac{1}{\epsilon T} \cr 
\tau^{-1} & -\tau^{-1} - D_s k^2 & 0 & 0 & \tau^{-1} \cr
\alpha\epsilon - \frac{\rho}{\tau} - \frac{1}{T} & \alpha\epsilon - 
\epsilon\nu - \frac{1}{T} & -2\alpha\epsilon & \epsilon\nu & 0 \cr 
-\frac{\sigma\epsilon}{1 + \epsilon} & \sigma & -\frac{\sigma}
{1 + \epsilon} & -\sigma - D_r k^2 & 0 \cr
-\frac{\sigma '\epsilon}{1 + \epsilon} & \sigma ' & -\frac{\sigma '}
{1 + \epsilon} & 0 & -\sigma ' - D_h k^2}\qquad 
\end{equation}
It is easy to study the behaviour of the eigenvalues of this matrix as a 
function of the parameters. For a broad range of parameters one finds three 
negative eigenvalues, one of which is very negative as found in the Smolin-
model \cite{1} stemming from the very large $D_r$. There are also two 
positive eigenvalues, the largest of these governs the evolution of the 
dominant instability of the disc\footnote{ For every set of parameters 
there are values of $k$ which give complex 
eigenvalues for $\lambda$ simultaneously above and below the real axis. It 
seems as if the system undergoes a phase transition but this has not been 
investigated. In the figures one can see the points beyond which
the eigenvalues become complex by looking for the kinks in the graphs above 
the $y$-axes. One notices that the real positive eigenvalues do not quite 
reach $0$ but this is of no consequence to the analysis since one can easily 
extrapolate from the point of the kink down to the $y$-axes due to the very 
regular behaviour of the largest positive eigenvalue with changing $k$. The 
reason why the figures were continued beyond the real positive eigenvalues is
a purely technical one and has to do with the difficulty of manipulating large
tables of numbers under Mathematica.}. 
In Fig.(1-2) the largest positive 
eigenvalue is plotted as a function of $y$, where $y = log(k)$, in units of 
$L_{int}$ i.e. $y=0$ corresponds to for example $L_{int}$, hence the largest scale 
are to the left towards the vertical axis and the smallest scales are to the 
right. 
\begin{figure}
\vskip-1in
\cfig{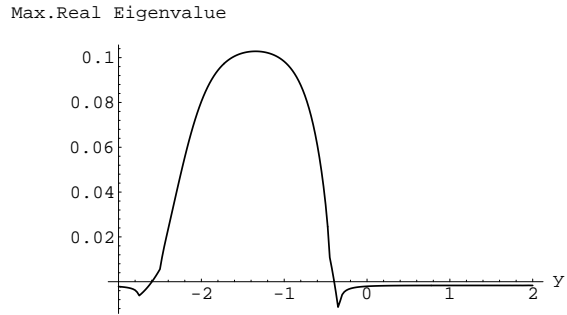}{4in}
\vskip-1in
\caption{$\nu/\alpha = 3,\ \ D_r/\sigma = 10^5$, $D_h/\sigma ' = 10$ 
and $D_s = 0$.}
\end{figure}
\begin{figure}
\vskip-1.25in
\cfig{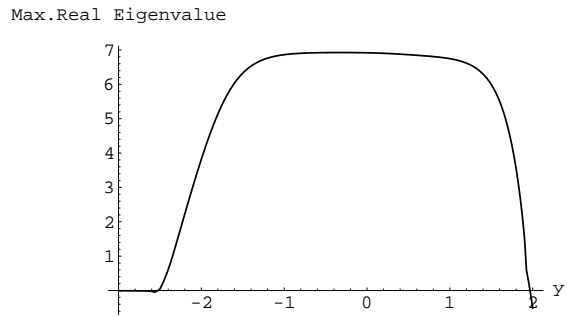}{4in}
\vskip-1in
\caption{$\nu/\alpha = 30$, $D_r/\sigma = 10^4$, such that the unstable mode 
does not run out of the graph (clearly unphysical situation) and 
$D_h/\sigma '$ as in Fig.1 with $D_s = 10^{-3}$.}
\end{figure}
\begin{figure}
\vskip-1in
\cfig{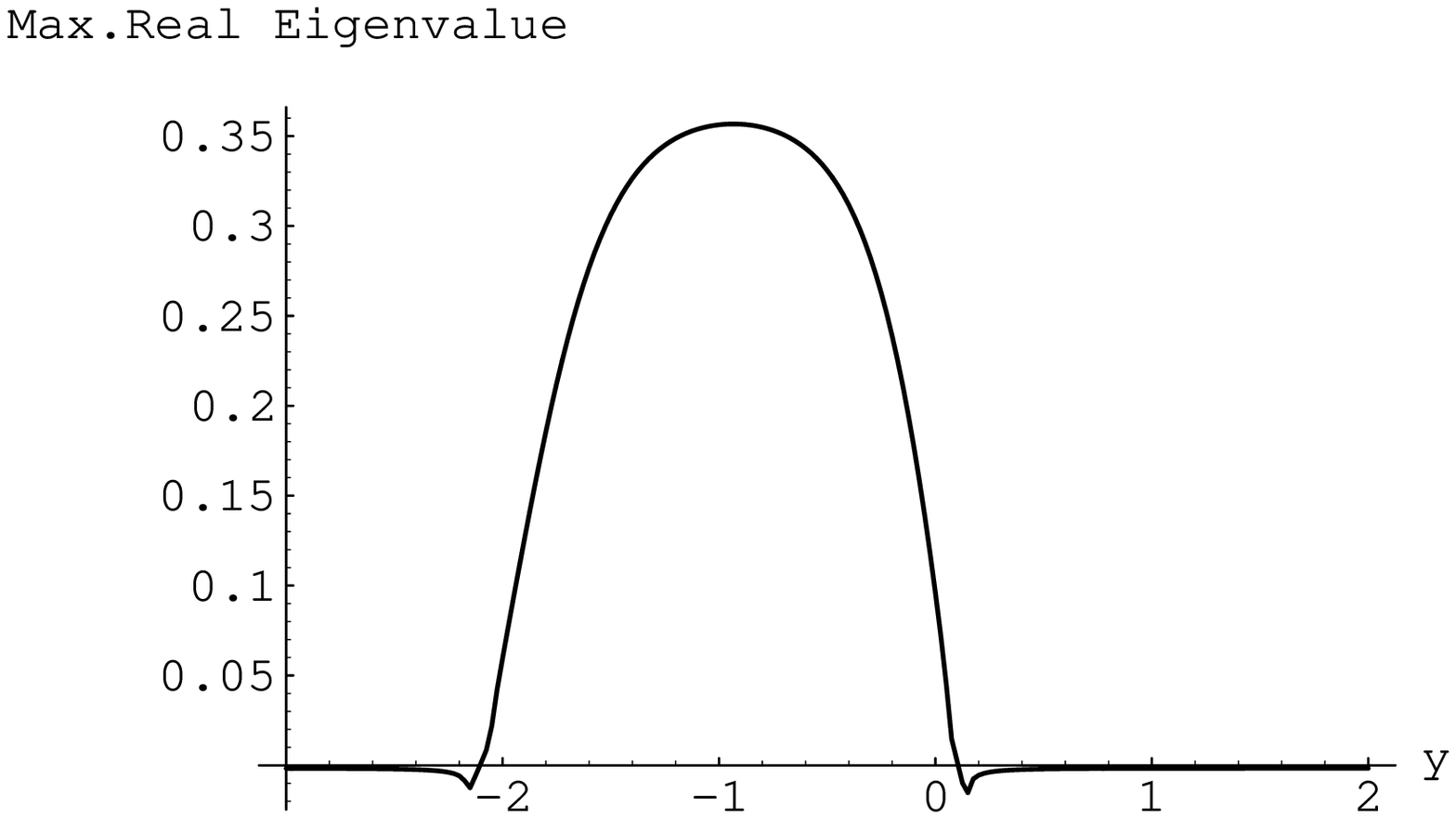}{4in}
\vskip-1.25in
\end{figure}
\begin{figure}
\vskip-1.25in
\cfig{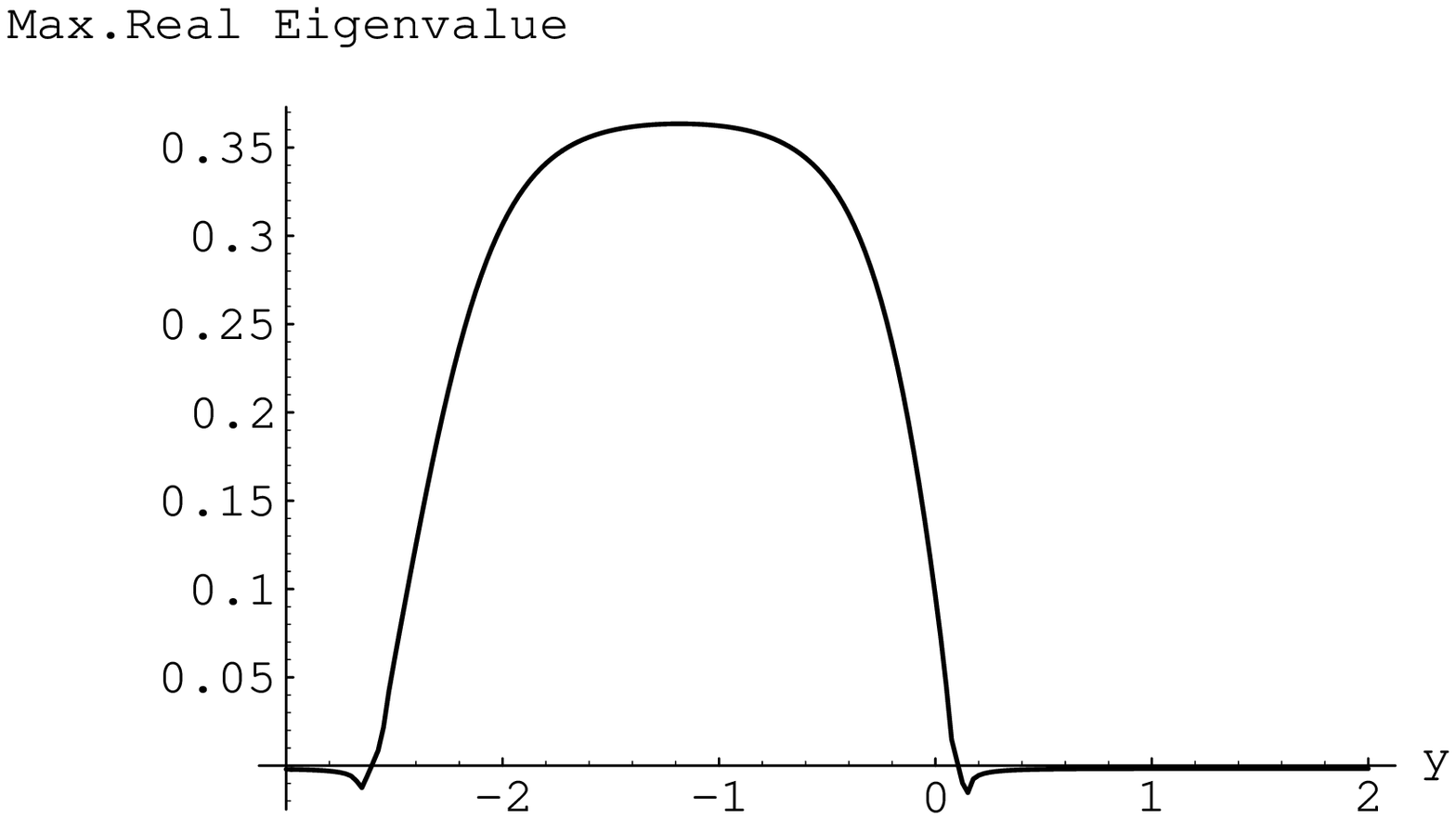}{4in}
\vskip-1.25in
\end{figure}
\begin{figure}
\vskip-1.25in
\cfig{gfig3c.eps}{4in}
\vskip-1in
\end{figure}
\begin{figure}
\vskip-1in
\cfig{gfig1.eps}{4in}
\vskip-1in
\caption{a) $D_r/\sigma = 10^5$  b) $D_r/\sigma = 10^4$ c) $D_h/\sigma ' = 1$ 
d) $D_h/\sigma ' = 10$. Note $D_s = 0$ in Fig.3(a-d)}
\end{figure} 
One observes by changing the parameters of the model that the most unstable 
modes at large scales are very stable towards parameter changes, however 
this is not true for small scales.

The time scale of the instability is between $10^6$ and $10^8$ years as can
be esaily deduced from the figures.
Even though at this 
stage the scales at which the instabilities occur seem to be right for the 
initiation of spiral structure formation as in Sc-type galaxies, the full non-
linear analysis with rotation has to be carried out to be sure whether this 
model indeed works. As was observed above, the short and long distance 
suppresion of 
instabilities is sensitive to the choice of $\sigma$, $\sigma '$, $\nu$,
$D_r$, $D_s$, $D_h$ and, amazingly enough, also to $\alpha$. In the following,
the parameters, not yet too constrained by observations, are varied 
independently from one another up and down by $1$ order of magnitude and 
the following conclusions can be drawn about the size of the parameter 
space allowing
instabilities to form and the importance of the different parameters (see 
Fig.(3)):
\begin{itemize}

\item The formation of instabilities and the length scales at which 
      they occur, do 
      not depend on the parameters separately but rather on the ratios 
      $\nu/\alpha$, $D_r/\sigma$ and $D_h/\sigma '$. The time scale
      depends strongly on $\nu/\alpha$ and much less strongly on the other
      ratios.

\item Since $D_s$ is at most $1$\footnote {Massive stars diffuse very slowly},
      one does not find any significant dependence
      of the model for $0\leq D_s \leq 1$ for $\nu/\alpha = 3$, hence one can 
      set it to $0$. For $\nu/\alpha = 30$, there is a strong dependence of 
      the short-distance scale instabilities on $D_s$. One recovers the same
      type of morphology as in Fig.2 for $D_s = 1$ but has no or only weak 
      suppression of short-distance scale instabilities for $D_s = 0$.

\item The ratio $\nu/\alpha$ strongly determines the onset of instabilities 
      as one obtains two different morphologies for e.g., $\nu/\alpha = 3$
      or $\nu/\alpha = 30$ (see Fig.(1-2)) at two different time scales. 
      This fact is not too surprising
      since $\alpha$ is the time of core collapse and $\nu$ determines the 
      influence of UV radiation on cloud condensation. Let us point out once
      more that for $\nu/\alpha = 3$, at a time scale of about $10^8$ years, 
      there is only one type of morphology
      \footnote{Unless one basically neglects the diffusion of shockwaves.},
      however for $\nu/\alpha = 30$, at a time scale of $10^6$ years, one 
      finds both morphologies from Fig.1,2 if one varies for example $D_s$.  

\item The ratio $D_h/\sigma '$ influences the scale of the onset of the
      instabilities more weakly (see Fig.(3c,d)). That the ratio influences
      the lower scale is not surprising since shockwaves act as an inhibitor
      on short scales and as a catalyst on intermediate scales.

\item The ratio $D_r/\sigma$ influences the maximum range of the instabilites, 
      but not too strongly (see Fig 3a,b). This again is not surprising 
      since UV radiation
      is supposed to inhibit star formation on large scales. However, it is 
      interesting to note that for a particluar ratio one obtains the typical
      size of a galaxy independent of what $D_r$ is\footnote{ $D_r$ is the 
      only place where the size of a typical galaxy enters. } as long as the
      ratio stays constant. Also note that $\sigma$ does not depend in any
      way on the size of a typical galaxy.

\item From the above and the parameter space search exemplified by 
      the figures, one can draw the conclusion that the region in parameter 
      space allowing for instabilities to occur on the right scales, is 
      fairly large, however not too large since varying the ratios by 2 
      orders of magnitude destroys the formation of unstable modes in the 
      model.

\end{itemize}   

\section{Conclusions}
\label{concl}

The conclusion is that, albeit only a linearized analysis
has been carried out, the model provides instabilities in time and space on 
the right scales and is not very arbitrary (parameter space region seems to 
be not too large). This gives rise to the hope that a full non-linear analysis
which is currently underway, will be successful and the model indeed contains
flocculent structures. Even if this is so, it is by no means correct to say 
that
this is indeed the way structure formation occurs in Sc-type galaxies. For this
statement to be made a renormalization group analysis, stage three of this 
project, will have to be carried out in order to make predictions for 
observations like the correlation function of for example cold clouds to cold 
clouds in Sc-type galaxies and these will then have to be confirmed by 
experiment before one can be confident that reaction-diffusion models work
for flocculent galaxies.

\section*{Acknowledgements}

I would like to take this opportunity to thank Lee Smolin, who interested and
educated me in this subject and with whom I had many fruitfull discussions. 
Furthermore, I would like to thank John Baker who, with his constructive 
criticism, benefitted the work tremendously. Also I would like to thank
Shoudan Liang and Peter Meszaros for their critical comments while this work
was in progress as well as Pablo Laguna and Jane Charlton who read over the 
paper and gave valuable suggestions for improvement.

\end{document}